\documentclass[pra,aps,reprint,superscriptaddress]{revtex4-1}
\usepackage{amsmath} 
\DeclareMathOperator{\Tr}{Tr}

\usepackage{amsthm} 
\usepackage{amssymb}	
\usepackage{graphicx} 

\usepackage[T1]{fontenc}
\usepackage{units}
\usepackage{bbold}
\usepackage{dsfont}
\usepackage{bm}

\renewcommand{\thesection}{\arabic{section}}

\begin{document}


\title{Imaging anomalous nematic order and strain in optimally doped BaFe$_2$(As,P)$_2$}

\author{Eric Thewalt}

\affiliation{Department of Physics, University of California, Berkeley, California 94720}
\affiliation{Materials Science Division, Lawrence Berkeley National Laboratory, Berkeley, California 94720}

\author{Ian~M.~Hayes}

\affiliation{Department of Physics, University of California, Berkeley, California 94720}
\affiliation{Materials Science Division, Lawrence Berkeley National Laboratory, Berkeley, California 94720}

\author{James~P.~Hinton}

\affiliation{Department of Physics, University of California, Berkeley, California 94720}
\affiliation{Materials Science Division, Lawrence Berkeley National Laboratory, Berkeley, California 94720}

\author{Arielle~Little}

\affiliation{Department of Physics, University of California, Berkeley, California 94720}
\affiliation{Materials Science Division, Lawrence Berkeley National Laboratory, Berkeley, California 94720}

\author{Shreyas~Patankar}

\affiliation{Department of Physics, University of California, Berkeley, California 94720}
\affiliation{Materials Science Division, Lawrence Berkeley National Laboratory, Berkeley, California 94720}

\author{Liang~Wu}

\affiliation{Department of Physics, University of California, Berkeley, California 94720}
\affiliation{Materials Science Division, Lawrence Berkeley National Laboratory, Berkeley, California 94720}

\author{Toni~Helm}

\affiliation{Department of Physics, University of California, Berkeley, California 94720}
\affiliation{Materials Science Division, Lawrence Berkeley National Laboratory, Berkeley, California 94720}

\author{Camelia~V.~Stan}

\affiliation{Advanced Light Source, Lawrence Berkeley National Laboratory, Berkeley, California 94720}

\author{Nobumichi~Tamura}

\affiliation{Advanced Light Source, Lawrence Berkeley National Laboratory, Berkeley, California 94720}

\author{James~G.~Analytis}

\affiliation{Department of Physics, University of California, Berkeley, California 94720}
\affiliation{Materials Science Division, Lawrence Berkeley National Laboratory, Berkeley, California 94720}

\author{Joseph~Orenstein}

\affiliation{Department of Physics, University of California, Berkeley, California 94720}
\affiliation{Materials Science Division, Lawrence Berkeley National Laboratory, Berkeley, California 94720}

\date{\today}

\begin{abstract}

    We present the strain and temperature dependence of an anomalous nematic phase in optimally doped BaFe$_2$(As,P)$_2$. Polarized ultrafast optical measurements reveal broken 4-fold rotational symmetry in a temperature range above~$T_c$ in which bulk probes do not detect a phase transition. Using ultrafast microscopy, we find that the magnitude and sign of this nematicity vary on a~${50{-}100}~\mu$m length scale, and the temperature at which it onsets ranges from~40~K near a domain boundary to~60~K deep within a domain. Scanning Laue microdiffraction maps of local strain at room temperature indicate that the nematic order appears most strongly in regions of weak, isotropic strain. These results indicate that nematic order arises in a genuine phase transition rather than by enhancement of local anisotropy by a strong nematic susceptibility. We interpret our results in the context of a proposed surface nematic phase.

\end{abstract}

\pacs{}

\maketitle


Iron-based superconductors~\cite{kamihara2008iron,takahashi2008superconductivity,rotter2008superconductivity} have been the subject of significant interest largely as a result of evidence for quantum criticality~\cite{shishido2010evolution, kasahara2010evolution, nakai2010unconventional, abrahams2011quantum, hashimoto2012sharp, walmsley2013quasiparticle, analytis2014transport, shibauchi2014quantum, kuo2016ubiquitous} accompanied by divergent nematic susceptibility~\cite{harriger2011nematic,yi2011symmetry,chu2012divergent,bohmer2013nematic,fernandes2014drives} in the vicinity of optimal doping.  These phenomena have been associated with an enhancement of the superconducting critical temperature~$T_c$~\cite{metlitski2015cooper,lederer2015enhancement,maier2014pairing}.

\begin{figure}[t]
  \includegraphics[width=3.375in]{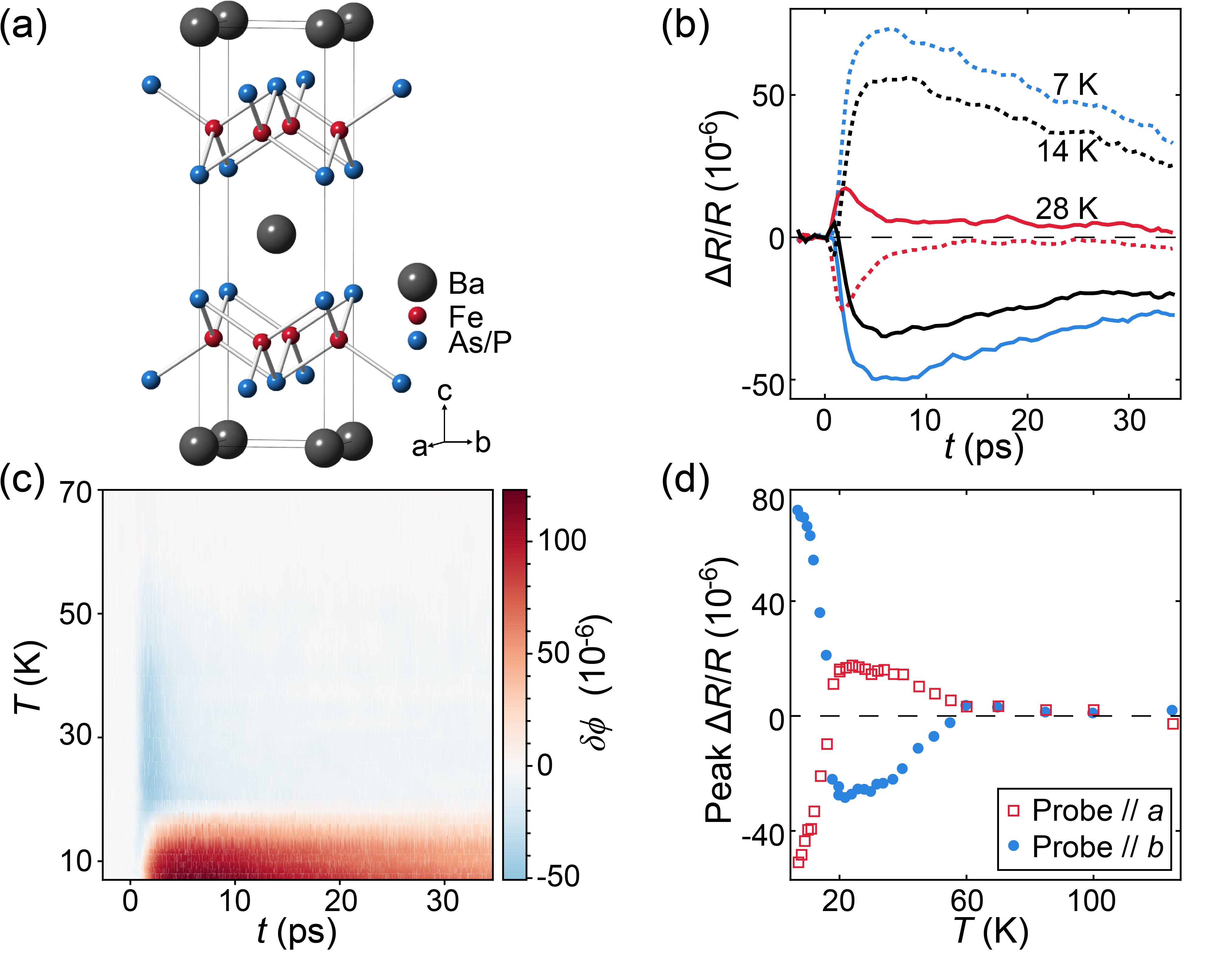}
  \caption{
      Crystal structure of P:Ba122 and photomodulation results at optimal doping.  (a)~Crystal structure of P:Ba122.  (b)~Pump/probe response $\Delta R/R$ as a function of time at a fixed position, with probe polarization parallel to the \mbox{Fe$-$Fe} directions~$a$ (solid) and~$b$ (dotted).  Red, black, and blue traces correspond to~${T=28}$~K, 14~K, and~7~K, spanning the apparent superconducting transition temperature.  (c)~Time and temperature dependence of the $C_4$-odd photomodulation response ${\delta \phi \equiv (\Delta R_b - \Delta R_a)/R}$. (d)~Temperature dependence of the maximum-amplitude value of~$\Delta R(t)/R$ for probe polarization along~$a$ (red) and~$b$ (blue), illustrating near-perfect antisymmetry under a~$\pi/2$ rotation of the probe polarization, abrupt onset of broken~$C_4$ symmetry, and competition between superconductivity and nematic order.
    \label{fg:t-T-struct}}
\end{figure}

Evidence for a quantum critical point (QCP) near optimal doping is particularly strong in BaFe$_2$(As$_{1-x}$P$_x$)$_2$, or P:Ba122, an isoelectronically doped superconductor.  At high temperature this material has a tetragonal crystal structure, shown in Fig.~\ref{fg:t-T-struct}(a), consisting of layers of Fe ions arranged in a square lattice with a pnictogen ion alternating above and below the center of each plaquette, and Ba ions between the layers.  The parent compound BaFe$_2$As$_2$ undergoes simultaneous tetragonal-to-orthorhombic and N\'{e}el spin-density-wave (SDW) transitions at~$T_N \approx 150~$K~\cite{si2016high}, breaking four-fold rotational~($C_4$) symmetry.  Substitution of As by P~\cite{allred2014coincident} and~$c$-axis compression~\cite{duncan2010high} each suppress~$T_N$ by reducing the average height of pnictogen ions and widening the Fe~$3d$ bands, which destabilizes the SDW order~\cite{rotter2010different}.  Bulk probes, including neutron and x-ray scattering, transport, NMR~\cite{hu2015structural}, and specific heat~\cite{walmsley2013quasiparticle}, indicate that the SDW phase onsets above~$T_c$ for P concentration up to, but not above,~${x=0.29}$, just below optimal doping~(${x=0.3}$).

Despite the evidence from these bulk probes, persistent hints that~$C_4$ symmetry is broken in samples with~$x>0.3$ suggest that there is more to the story. Angle-resolved photoemission (ARPES)~\cite{shimojima2014pseudogap,sonobe2015orbital} and torque magnetometry~\cite{kasahara2012electronic} studies have found evidence of broken~$C_4$ symmetry above the dome of superconductivity persisting above optimal doping in P:Ba122, and optical data suggest similar behavior in Ba(Fe,Co)$_2$As$_2$~\cite{stojchevska2012doping}.

The simplest explanation for this apparent discrepancy is that typical samples are under strain.  This strain can either be frozen in during crystal growth, which we call intrinsic strain, or caused by the crystal mounting and cooling processes, which we call extrinsic strain.  Such strain, when coupled with diverging nematic susceptibility near the QCP, would induce nematic order that would strengthen rapidly but smoothly with decreasing temperature. However, the measurements of nematicity at~$x>0.3$ indicate that it tends to have an abrupt onset~\cite{shimojima2014pseudogap,sonobe2015orbital,stojchevska2012doping}, and our results corroborate this observation.

In this letter we present a study of nematicity in optimally doped P:Ba122, with the aim of resolving the apparent contradiction between implications from different experiments. We map a single region of a P:Ba122 crystal with two local probes of broken~$C_4$: time-resolved optical pump/probe reflectance, or photomodulation, which enhances weak structure in the reflectance~$R$~\cite{cardona1967electroreflectance}; and scanning Laue microdiffraction~\cite{tamura2003scanning}, which allows us to explore the link between local strain and the onset and strength of nematicity.  Our photomodulation measurements reveal nematic order above~$T_c$, with magnitude, sign, and onset temperature varying on a length scale of~$50{-}100~\mu$m.

Contrary to expectation, we find that the nematic order observed via photomodulation is strongest in regions where uniaxial strain and transverse dilation are weakest.  However, the boundaries of domains of nematic order coincide with sharp features in local strain.  This suggests that the nematic order develops in a genuine phase transition rather than as a result of local anisotropy amplified by strong nematic susceptibility.  Our results are consistent with a surface nematic phase, as has been suggested by calculations incorporating interlayer hopping~\cite{song2016surface}.  The existence of such a phase would relieve the tension between results from bulk and surface probes.


Measurements of photomodulated reflectance,~$\Delta R$, were performed using linearly polarized, 100~fs-duration pulses from a mode-locked Ti:Sapphire laser at 80~MHz repetition rate, 800~nm center wavelength, and~${{\sim} 5~\mu\text{J}/\text{cm}^2}$ fluence.  Our initial measurements showed strong dependence of the amplitude and sign of~$\Delta R$ on the position of the pump/probe focus on the sample surface. As a result, local characterization of the time and temperature dependence of~$\Delta R$ required accurate stabilization of the position of the laser focus relative to the sample during cooling.  This was achieved by registering the sample to an optical landmark on its mount using a high-resolution video feed, enabling us to fix the focal position with a precision of 5~$\mu$m.  Figure~\ref{fg:t-T-struct}(b) shows examples of pump/probe traces measured at a fixed position on a sample with~$x=0.31$ at three temperatures spanning the apparent superconducting transition, with the probe polarized along the orthogonal \mbox{Fe$-$Fe} directions, which we (arbitrarily) label~$a$ and~$b$ (solid and dotted, respectively).  (The stated temperatures are nominal; the actual crystal temperature at the laser focus is higher as a result of laser heating.  We studied the apparent superconducting transition temperature as a function of laser fluence and confirmed that~$T_c$ approaches~31~K at low fluence; the data are presented in Sec.~\ref{sec:fluence-dependence}.)

The photomodulation data show striking evidence of broken~$C_4$ symmetry.  In the presence of~$C_4$ symmetry~$\Delta R$ would be independent of the polarization of the probe electric field; that is, ${\Delta R_a=\Delta R_b}$.  Instead, the pump/probe response is approximately equal and opposite along orthogonal \mbox{Fe$-$Fe} directions, i.e.~${\Delta R_a \approx -\Delta R_b}$.  In subsequent discussion we consider the strength of the $C_4$-odd component of the photomodulation response, ${(\Delta R_b - \Delta R_a)/R\equiv \delta \phi}$, to be a proxy for nematic order (see Sec.~\ref{sec:order-parameter} for details).

The full time and temperature dependence of~$\delta \phi$ is shown in Fig.~\ref{fg:t-T-struct}(c).  There are two distinct forms of pump/probe response: above the superconducting transition, the response is short-lived and~$\delta \phi$ is negative; well below $T_c$, the response is long-lived and~$\delta \phi$ is positive.  Near the transition, both forms are apparent.  To better illustrate the singular features of the temperature dependence, we plot in Fig.~\ref{fg:t-T-struct}(d) the maximum-amplitude value of~$\Delta R(t)/R$ as a function of temperature for~$a$ and~$b$ probe polarizations.  With decreasing temperature,~$\Delta R$ first appears abruptly above the noise at~${\sim} 60$~K.  Upon further cooling, the sign of~$\Delta R$ changes abruptly near~$T_c$, and at low temperature the sign is reversed relative to the normal state.

The change in sign and relaxation rate at~$T_c$ can be understood on the basis of competition between the nematic order parameter,~$\phi$, and the superconducting order parameter,~$\psi$.  For~${T > T_c}$, the pump pulse weakens the nematic order, which then returns rapidly to its equilibrium value.  However, for~${T < T_c}$ the pump also suppresses~$\psi$, and since the timescale of this suppression is longer than that of the nematic order a quasiequilibrium results in which~$\phi$ is enhanced due to the mutual repulsion of~$\phi$ and~$\psi$.  The enhancement of~$\phi$ persists until~$\psi$ returns to its equilibrium amplitude.  Section~\ref{sec:only-nematic} contains a detailed discussion of this model.


The observation of broken~$C_4$ at a fixed location on the sample surface strongly suggests domain formation as the origin of the position dependence described above.  To test this hypothesis, we mapped the variation of~$\delta \phi$ on the sample surface. These maps were obtained by mounting samples onto an~$xyz$ piezo-stage, and scanning the sample with respect to an 8~$\mu$m diameter focus of overlapping pump and probe beams.  The P:Ba122 crystal was mounted on a Cu plate, providing a net 0.2\% compressive strain on the base of the sample via thermal contraction.

\begin{figure}[t]
  \centering
  \includegraphics[width=3.375in]{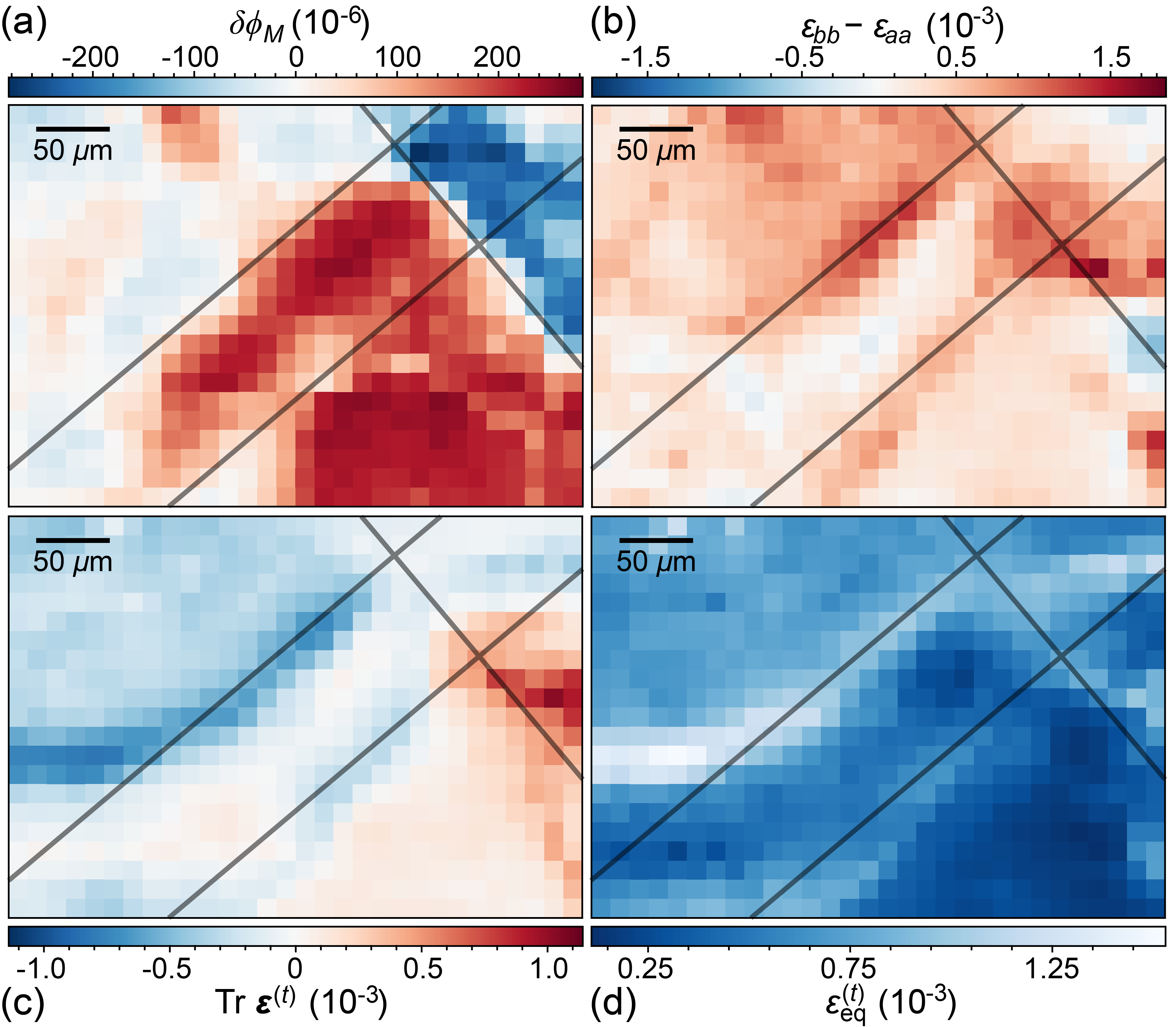}
  \caption{
    Spatial variation (13~$\mu$m resolution) of optical anisotropy~(a) and~$ab$-plane strain~(b-d) on a~$390\times 260~\mu$m region of an optimally doped P:Ba122 crystal mounted on~Cu.  (a)~Photomodulation proxy for nematic order,~$\delta\phi_M$.  (b)~Transverse strain anisotropy~$\varepsilon_{bb}-\varepsilon_{aa}$ in the \mbox{Fe$-$Fe} basis.  (c)~Transverse unit cell dilation~$\Tr \bm{\varepsilon}^{(t)}$.  (d)~Transverse equivalent strain $\varepsilon^{(t)}_\text{eq}=(2\varepsilon^{(t)}_{ij}\varepsilon^{(t)}_{ij}/3)^{1/2}$.  Superimposed lines are parallel to the \mbox{Fe$-$Fe} directions and are located at the same positions in each image to facilitate visual comparison of features.  Optical data were collected at~$T=5$~K; strain data at room temperature. 
    \label{fg:maps}}
\end{figure}

A map of local nematicity obtained by spatially resolved photomodulation is shown in Fig.~\ref{fg:maps}(a). The color of each square encodes the maximum-amplitude value,~$\delta\phi_M$, of~${(\Delta R_b(t) - \Delta R_a(t))/R}$; that is, of the difference between~$\Delta R$ measured along the two principal axes. Domain boundaries separating regions of broken~$C_4$ symmetry with orthogonal nematic order are readily apparent.  We note that the typical domain size of~${\sim}100~\mu$m is approximately the size of crystals used in the previously cited torque magnetometry experiments that suggested a broad nematic phase above the superconducting dome~\cite{kasahara2012electronic}.

The spatial patterns of positive and negative~$\delta\phi_M$ do not change with repeated heating and cooling of the sample, suggesting that the magnitude and sign of the nematic order are determined by some local quantity.  A local strain field, perhaps frozen into the crystal during growth, is a natural candidate; a difference between the strains along orthogonal \mbox{Fe$-$Fe} directions would couple directly to~$C_4$-breaking order~\cite{kuo2012magnetoelastically}.  Another potential contributing factor is local in-plane compression of the unit cell~\cite{bohmer2017effect}, which would increase the pnictogen height and the \mbox{Fe$-$As$-$Fe} bond angle, counteracting the effect of P doping~\cite{rotter2010different} and driving the crystal back toward the underdoped SDW phase.


In order to explore the link between local strain and the onset of nematic order, we used scanning Laue (i.e., polychromatic) microdiffraction to map the local strain at room temperature in the same region of the sample that was imaged using photomodulation (see Sec.~\ref{sec:region-transform} for details on the region-alignment procedure).  A full diffraction pattern was collected at each position and used, along with the known lattice parameters, to extract the deviatoric (i.e., traceless) strain tensor~$\bm{\varepsilon}$, which describes the local deformation of the unit cell.  In a given basis, the diagonal components~$\varepsilon_{aa}$,~$\varepsilon_{bb}$, and~$\varepsilon_{cc}$ of the strain tensor correspond to expansion (or compression, for negative values) along the corresponding direction, while the off-diagonal components~$\varepsilon_{ab}$,~$\varepsilon_{bc}$, and~$\varepsilon_{ca}$ correspond to pure shear.  Since we are primarily concerned with strain in the \mbox{Fe$-$As} layers, we focus on the~$ab$ subsector of~$\bm{\varepsilon}$, which we denote by~$\bm{\varepsilon}^{(t)}$.  The dilation of the~$ab$-plane unit cell is given by~$\Tr \bm{\varepsilon}^{(t)}=\varepsilon_{aa}+\varepsilon_{bb}$; compression corresponds to negative values.

Figure~\ref{fg:maps} illustrates the relationship between the previously discussed map of low-temperature optical anisotropy in Fig.~\ref{fg:maps}(a) and the spatial variation of the strain tensor in Figs.~\ref{fg:maps}(b-d).  The superimposed lines, oriented with the \mbox{Fe$-$Fe} directions~$a$ and~$b$, are positioned identically on each image.  Figure~\ref{fg:maps}(b) shows the strain anisotropy in the \mbox{Fe$-$Fe} basis,~${\varepsilon_{bb}-\varepsilon_{aa}}$, in the same region of the crystal.  Contrary to what would be expected if the nematic order were the result of a local strain bias, the changes in sign of~$\delta \phi_M$ and the \mbox{Fe$-$Fe} strain anisotropy do not coincide.  Furthermore, the \mbox{Fe$-$Fe} strain anisotropy is small in magnitude in most of the region where the nematic photomodulation response is strongest.  Figure~\ref{fg:maps}(c) shows the transverse unit-cell dilation~$\Tr \bm{\varepsilon}^{(t)}$, which is small and mostly positive in the large region corresponding to large positive~$\delta \phi_M$, contradicting the prediction that negative~$\Tr \bm{\varepsilon}^{(t)}$ would drive the system toward the~$C_4$-breaking SDW phase.  Finally, Fig.~\ref{fg:maps}(d) shows the equivalent strain~${\varepsilon^{(t)}_\text{eq}=(2\varepsilon^{(t)}_{ij}\varepsilon^{(t)}_{ij}/3)^{1/2}}$, a measure of total strain.  Although the nematic order and the strain anisotropy are not strongly correlated, the edges of the nematic domains are coincident with strain features; in particular, with local maxima in the equivalent strain and with extrema in~$\Tr \bm{\varepsilon}^{(t)}$.  (We note that the observed strain variations are likely intrinsic rather than extrinsic, as we observed similar variations in an optimally doped crystal mounted strain-free; see~\ref{sec:strain-free-mount} for details.)

Taken together these results strongly suggest that local strain is not the driver, via divergent susceptibility, of the nematicity we observe -- in fact, strong strain anisotropy (and strong strain in general) appears to suppress the electronic nematicity.

\begin{figure}[t!]
  \centering
  \includegraphics[width=3.375in]{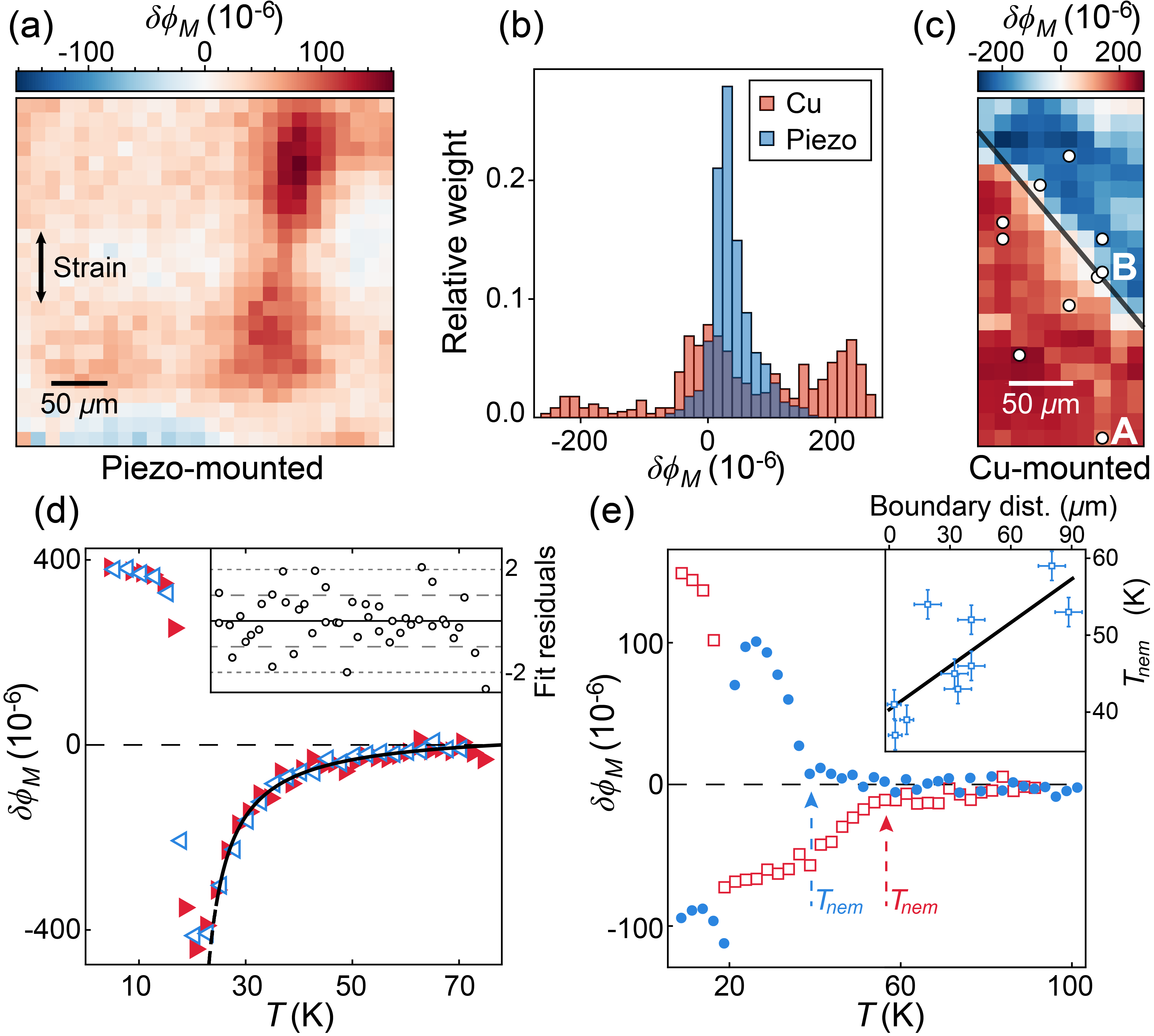}
  \caption{
    Comparison of spatial variation (13~$\mu$m resolution) and temperature dependence of nematic order for piezo-mounted (uniaxially strained) and Cu-mounted (biaxially strained) crystals.  (a)~Spatial variation of photomodulation proxy for nematic order,~$\delta\phi_M$, on the piezo-mounted crystal, which is uniaxially strained as indicated.  (b)~Histograms showing distribution of~$\delta\phi_M$ for both crystals.  (c)~Spatial variation of~$\delta\phi_M$ on the Cu-mounted crystal, with open circles indicating positions at which temperature dependence data was collected and black line marking a region of null~$\Delta R/R$ response separating regions of opposite nematic sign.  (d)~Temperature dependence of~$\delta\phi_M$ for the piezo-mounted crystal while warming (right-pointed triangles) and cooling (left-pointed triangles).  The black line is a Curie-Weiss fit with~$T_{CW}=19~$K (solid on fitted domain; dashed at lower temperatures).  Inset:~standardized fit residuals.  (e)~Temperature dependence of~$\delta\phi_M$ for the Cu-mounted crystal far from the boundary at the point marked~A (open squares) and near the boundary at the point marked~B (circles).  Apparent nematic transition temperatures are indicated.  Inset:~scatter plot of nematic transition temperature and distance from the domain boundary indicated by the black line in (c); correlation is positive with $p$-value~$< 10^{-2}$.
  \label{fg:maps-temps-histo}}
\end{figure}

In order to further study the effect of extrinsic uniaxial strain, we also performed ultrafast microscopy on an optimally doped sample mounted on a piezoelectric stack.  On cooling, the piezo provides a tensile uniaxial strain by thermally contracting by 0.1\% (similar to optimally doped P:Ba122) along one lateral dimension while expanding by 0.1\% along the other.  The crystal's \mbox{Fe$-$Fe} directions were aligned with these principal piezo axes.  The resulting image of~$\delta \phi_M$ is shown in Fig.~\ref{fg:maps-temps-histo}(a).  The domain population of the uniaxially strained crystal differs significantly from that of the Cu-mounted sample, as is evident in Fig.~\ref{fg:maps-temps-histo}(b), which compares histograms of~$\delta\phi_M$ in both samples.  The uniaxial strain appears to bias the domain population, shifting the central Cu peak while suppressing the large-amplitude nematic response.  Thus, while intrinsic strain defies expectation, extrinsic strain biases the electronic nematicity in the expected manner.

In addition to pump-probe microscopy, we measured the temperature dependence of~$\delta\phi_M$ on both crystals, including at multiple points on the Cu-mounted sample.  These points are indicated by white circles in Fig.~\ref{fg:maps-temps-histo}(c), and the points marked~A and~B correspond respectively to the red and blue~$\delta \phi_M(T)$ markers in Fig.~\ref{fg:maps-temps-histo}(e), where~$\delta \phi_M$ is plotted as a function of temperature.

The onset of the nematic optical response in the Cu-mounted crystal is abrupt and manifestly non-Curie-Weiss, and the temperature at which it onsets varies between approximately~40~K and~60~K depending on position, as illustrated in Fig.~\ref{fg:maps-temps-histo}(e).  We found that~$T_\text{nem}$ is positively correlated (with $p$-value $< 10^{-2}$) with distance from the boundary of null nematic response indicated by a black line in Fig.~\ref{fg:maps-temps-histo}(c); the data and fit are shown in the inset of Fig.~\ref{fg:maps-temps-histo}(e).  We observed no statistically significant correlation between superconducting and nematic onset temperatures, indicating that the effects we see are not a result of inhomogeneity in P~concentration.

In contrast to the Cu-mounted sample, the temperature dependence of~$\delta \phi_M$ on the piezo-mounted crystal is well-described by a Curie-Weiss form with transition temperature~${T_{CW}=19}$~K.  The fit (black line; solid on fitted region) and data are shown in Fig.~\ref{fg:maps-temps-histo}(d), with the standardized fit residuals in the inset.  In the presence of strong, uniform uniaxial strain, therefore, we observe a nematic onset that is consistent with the picture of divergent nematic susceptibility, which makes the sharpness of the nematic onset in the Cu-mounted sample all the more notable.  We do not observe any hysteretic difference between the data collected with increasing temperature (right-pointed markers) and with decreasing temperature (left-pointed markers).

The strong correlation between the nematic onset temperature and distance from the boundary between the positive and negative domains suggests that we may be observing a nucleation phenomenon, where the nematic domains arise deterministically at some distant crystalline features and then spread as the temperature decreases until they reach the high-equivalent-strain boundaries indicated in Fig.~\ref{fg:maps}(d).  This picture is particularly compelling in light of recent work incorporating hopping between \mbox{Fe$-$As} layers, which has shown that interlayer hopping can produce a surface nematic phase that onsets at significantly higher temperatures than in the bulk~\cite{song2016surface}.  A surface phase, which could also arise due to stabilization of fluctuating order by soft surface phonons~\cite{brown2005surface}, would be more susceptible to confinement by boundaries of strain due to the reduced dimensionality and volume of the required region of contiguous deformation, and could be disfavored under transverse compression due to buckling-induced disorder.  In addition, this model is consistent both with surface measurements that indicate a genuine nematic phase~(\cite{shimojima2014pseudogap,sonobe2015orbital,stojchevska2012doping}, this work) and with bulk measurements that show no evidence of a phase transition~\cite{hu2015structural, walmsley2013quasiparticle, allred2014coincident}.


In conclusion, photomodulation measurements reveal that optimally doped BaFe$_2$(As,P)$_2$ has a~$C_4$-breaking phase well above~$T_c$ that varies strongly in magnitude, sign, and onset temperature at length scales of~$50{-}100~\mu$m.  Scanning Laue microdiffraction measurements show that the local strain anisotropy and local transverse compression of the unit cell, which are both expected to favor nematic order, are anticorrelated with the observed optical nematicity.  These results imply that the optical nematicity in the biaxially strained crystal corresponds to a genuine nematic phase transition rather than amplification of local anisotropy by enhanced nematic susceptibilty.  We interpret this phase as a surface phenomenon~\cite{song2016surface} that nucleates well above~$T_c$ and spreads until it reaches boundaries where the crystal is highly strained.  A surface nematic phase with large domains reconciles ARPES~\cite{shimojima2014pseudogap,sonobe2015orbital}, optical~\cite{stojchevska2012doping}, and torque magnetometry~\cite{kasahara2012electronic} measurements showing nematic order at optimal doping with bulk measurements~\cite{hu2015structural,walmsley2013quasiparticle,allred2014coincident} that do not show a phase transition.  In general, phase diagrams of two-dimensional materials may differ significantly from those based on bulk measurements of the same compound.

\begin{acknowledgments}
We thank E. Angelino, R. Fernandes, I. Fisher, F. Flicker, A. Koshelev, K. Song, and N. Yao for helpful discussions.  Measurements and modeling were performed at the Lawrence Berkeley National Laboratory in the Quantum Materials program supported by the Director, Office of Science, Office of Basic Energy Sciences, Materials Sciences and Engineering Division, of the U.S. Department of Energy under Contract No.  DE-AC02-05CH11231.  Synthesis of P:Ba122 was supported by Laboratory Directed Research and Development Program of Lawrence Berkeley National Laboratory under Contract No. DE-AC02-05CH11231.  J.O., L.W., and A.L. received support for performing and analyzing optical measurements from the Gordon and Betty Moore Foundation's EPiQS Initiative through Grant GBMF4537 to J.O. at UC Berkeley.  Material synthesis and characterization was supported by the Gordon and Betty Moore Foundation's EPiQS Initiative Grant GBMF4374 to J.A. at UC Berkeley.  Laue microdiffraction measurements were carried out at beamline 12.3.2 at the Advanced Light Source.  The ALS is supported by the Director, Office of Science, Office of Basic Energy Sciences, of the U.S.  Department of Energy under Contract No. DE-AC02-05CH11231.
\end{acknowledgments}

\bibliography{biblio}


\clearpage
\begin{center}
\textbf{\large Supplemental Materials: \\ Imaging anomalous nematic order and strain in optimally doped BaFe$_2$(As,P)$_2$}
\end{center}

\setcounter{equation}{0}
\setcounter{figure}{0}
\setcounter{table}{0}
\setcounter{page}{1}
\setcounter{section}{0}
\makeatletter
\renewcommand{\theequation}{S\arabic{equation}}
\renewcommand{\thefigure}{S\arabic{figure}}
\renewcommand{\thesection}{S\arabic{section}}
\renewcommand{\bibnumfmt}[1]{[S#1]}
\renewcommand{\citenumfont}[1]{S#1}

\section{\label{sec:fluence-dependence}Fluence dependence of~$T_c$}

\begin{figure}[h!]
  \centering
  \includegraphics[width=3.375in]{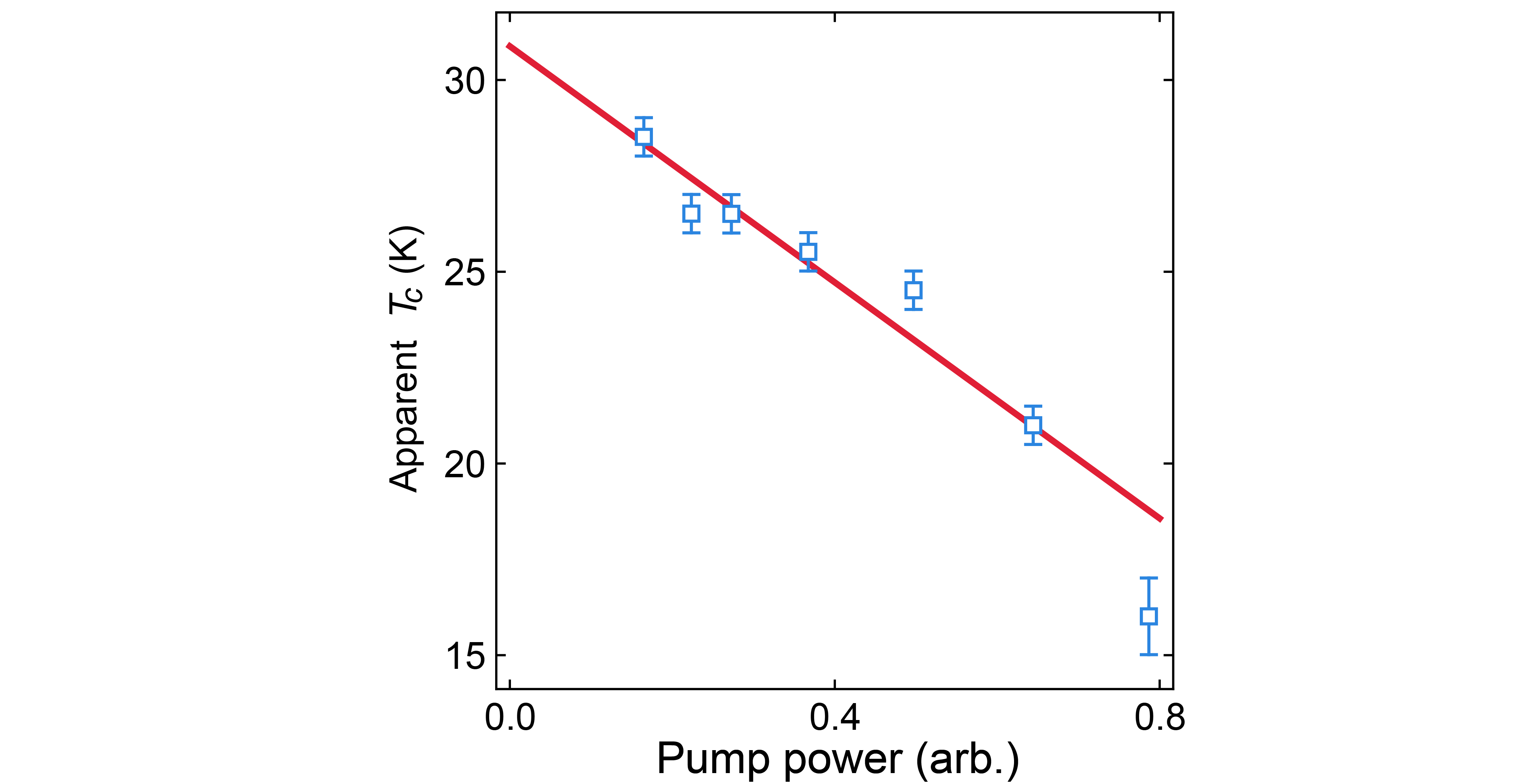}
  \caption{Apparent superconducting transition temperature~$T_c$ as a function of pump power in an optimally doped P:Ba122 sample.  At zero power, the extrapolated~$T_c$ is~$30.9\pm0.9$~K.
  \label{fg:tc-fludep}}
\end{figure}

We collected photomodulation data as a function of temperature for a range of pump fluences.  Figure~\ref{fg:tc-fludep} shows the apparent superconducting transition temperature~$T_c$ as a function of pump power.  While pump-induced heating has a significant effect on nominal~$T_c$ at high fluence, a linear regression yields a low-power limit of~${T_c=30.9\pm0.9}$~K, consistent with expectations.

\section{\label{sec:order-parameter}Photomodulation anisotropy as a proxy for~$C_4$-breaking order}

\begin{figure}[h!]
  \centering
  \includegraphics[width=3.375in]{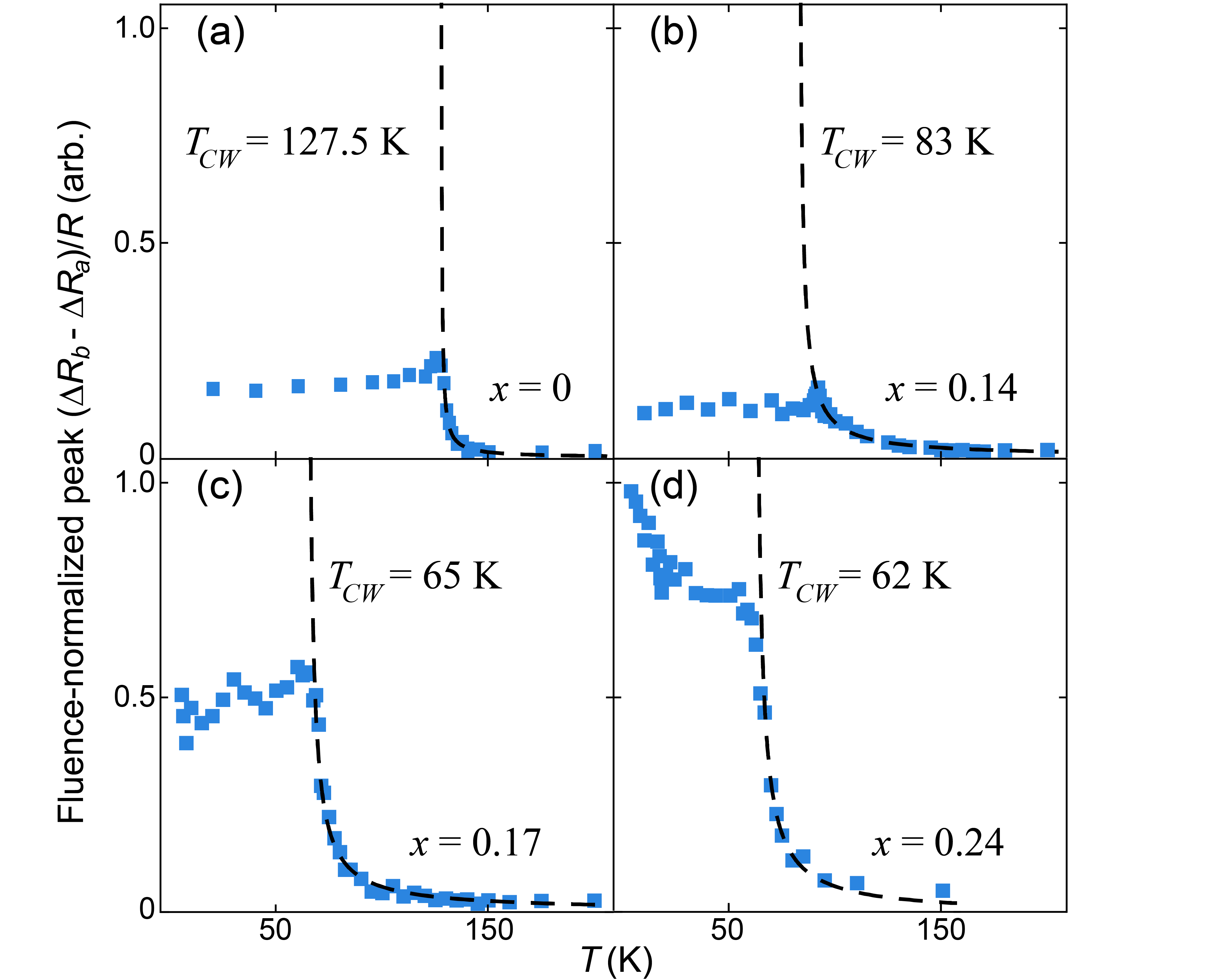}
    \caption{Normalized peak antisymmetrized pump/probe response,~${(\Delta R_b - \Delta R_a)/R}$, as a function of temperature for underdoped P:Ba122 samples with P concentration~(a)~${x=0}$, (b)~$0.14$, (c)~$0.17$, and (d)~$0.24$.  Fits to a Curie-Weiss form proportional to~$1/(T-T_{CW})$ are indicated by dashed lines.
  \label{fg:drr-order-param}}
\end{figure}

In the pump/probe reflectivity experiment, the pump pulse, which generically causes an incoherent excitation of electrons, can couple to an order parameter~$\Phi$ in two characteristic ways: 1) by raising the electron temperature~$T_e$, in which case ${\Delta R \propto \delta \Phi/\delta T_e}$ and~$\Delta R$ will be peaked near its transition temperature; or 2) by directly suppressing the order by depopulating relevant electronic states, in which case $\Delta R \propto \Phi$.

We performed a detailed study of the temperature dependence of the nematic pump/probe response in underdoped P:Ba122.  In Fig.~\ref{fg:drr-order-param} we plot the peak (maximum amplitude) value of~${(\Delta R_b - \Delta R_a)/R}$ as a function of temperature for four underdoped P:Ba122 samples, along with fits to a Curie-Weiss temperature dependence.  This quantity, which is odd under a~$\pi/2$ rotation, has the symmetry of a nematic order parameter, and is approximately proportional to the~$C_4$-breaking order parameter rather than its temperature derivative.  This indicates that~${(\Delta R_b - \Delta R_a)/R}$ can be used as a proxy for nematic order in the P:Ba122 system.

\section{\label{sec:only-nematic}Competition between superconducting and nematic order}

Whatever the origin of $C_4$-breaking order at~${x=0.31}$, the temperature dependence of the photomodulation~(PM) amplitude shown in Figs.~\ref{fg:t-T-struct}(b-d) and~\ref{fg:maps-temps-histo}(d-e) indicates that the nematic order is strongly coupled with superconductivity.  In this section we discuss the implications of the sign change in~${\delta\phi \equiv (\Delta R_b - \Delta R_a)/R}$ near~$T_c$, along with the fact that we generically observe proportionality $\Delta R_a(T) \propto \Delta R_b(T)$ at optimal doping.

In the presence of superconducting order, represented by~$\psi$, and a~$C_4$-breaking order~$\phi$, the pump/probe response includes contributions from both,

\begin{equation}
    \begin{pmatrix}
        \Delta R_a \\
        \Delta R_b
    \end{pmatrix}=\left(\alpha_e \bm{1} + \alpha_o \bm{\sigma}_z\right)\phi+\left( \beta_e \bm{1} + \beta_o \bm{\sigma}_z\right) \left|\psi \right|^2,
    \label{eq:pump-probe}
\end{equation}

\noindent where~$\bm{1}$ and~$\bm{\sigma}_z$ are the unit and Pauli $z$ matrices, respectively, and~$\alpha_e$ and~$\alpha_o$~($\beta_e$ and~$\beta_o$) are coefficients describing the even and odd components, respectively, of~$\Delta R$ under a~$\pi/2$ rotation of probe polarization due to the pump-induced weakening of the~$\phi$~($\psi$) order.  Equation~\ref{eq:pump-probe} implies that the observed proportionality of~$\Delta R_{a,b}$ for all~$T$ requires ${\alpha_e/\alpha_o=\beta_e/\beta_o}$; that is, that the ratio of symmetric and antisymmetric PM coefficients is the same for both forms of order.  The microscopic origin of reflectivity modulation induced by distinct orders will differ, so this equality would be a significant coincidence.  We hypothesize instead that the direct contribution of the pump-induced modulation of~$\psi$ is too small to observe; i.e., ${\beta_e\approx\beta_o\approx 0}$, so that even below~$T_c$ the pump/probe response arises exclusively from~$\phi$, and $\psi$ is visible due to a repulsive interaction between the two phases of the form~$\lambda \phi^2 |\psi|^2$ in the Ginzburg-Landau free energy.

Above the superconducting transition, where~${\psi=0}$,~${\phi\neq0}$, the PM response can be described by the rate equation ${\delta_t \phi=G(t)-\gamma(\phi-\phi_0)}$, where~$G$ describes the pump-induced weakening of~$\phi$ and~$\gamma$ is the rate at which it returns to its equilibrium value,~$\phi_0$.  We assume that the system is in a linear regime where~$G$ is proportional to the laser fluence;~$G$ is also proportional to~$\phi_0$, as established in Sec.~\ref{sec:order-parameter}.  In the short-pulse approximation, ${G(t)=-F\phi_0 \delta(t)}$, where~$F$ is proportional to the number of photons in the pump pulse.  Integrating the rate equation yields ${\phi(t)=[1-F \exp(-\gamma t)]\phi_0(T)}$ for~$t>0$. 

\begin{figure}[t!]
  \centering
  \includegraphics[width=3.375in]{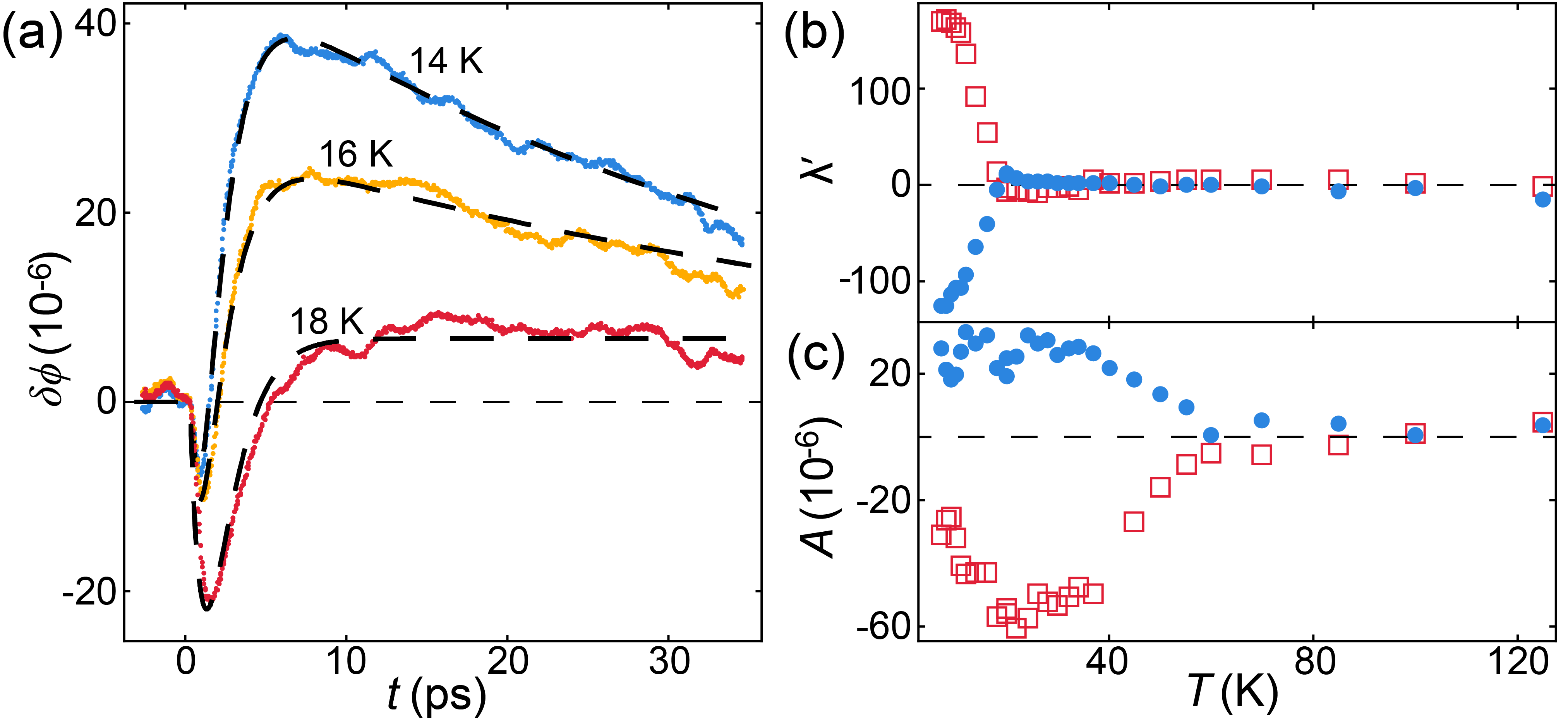}
    \caption{Fit results from a model wherein the suppression of the superconducting order parameter appears in~${\delta \phi \equiv (\Delta R_b - \Delta R_a)/R}$ only through the enhancement it induces in the nematic order parameter.  (a)~Example fits (dashed black) to the traces with ${T=14~K}$, 16~K, and~18~K.  (b)~Temperature dependence of the fit parameter~$\lambda'$ corresponding to the amplitude of the superconducting response, with probe polarization parallel to the~$a$ axis (blue circles) and~$b$ axis (red squares).  (c)~Temperature dependence of the fit parameter~$A$ corresponding to the amplitude of the nematic response, with probe polarization parallel to the~$a$ axis (blue circles) and~$b$ axis (red squares). 
  \label{fg:nematic-fit}}
\end{figure}

In the superconducting phase, the repulsive interaction energy~$\lambda\phi^2 |\psi|^2$ reduces the equilibrium order to ${\phi_0(T)=\phi^*_0(T)(1-2\chi_\phi \lambda |\psi_0(T)|^2)^{-1}}$, where~$\phi^*_0$ is the order parameter amplitude in the absence of coupling to superconductivity,~$\psi_0$ is the equilibrium value of the superconducting order parameter, and~$\chi_\phi$ is the nematic susceptibility.  We assume that in the superconducting phase the pump pulse weakens both~$\phi$ and~$|\psi|^2$.  Following photoexcitation,~$\phi$ recovers not towards~$\phi_0$, but rather to a time-dependent quasi-equilibrium value~$\phi_1(t)$, given by

\begin{equation*}
    \phi_1(t)\approx \phi_0\left(1+\frac{2\lambda\chi_\phi\delta|\psi(t)|^2}{1-2\lambda\chi_\phi|\psi(t)|^2}\right),
\end{equation*}

\noindent which exceeds~$\phi_0$ because of the pump-induced weakening of the superconducting order,~$\delta|\psi|^2$.  In the presence of dynamic coupling,~$\phi(t)$ will decay towards~$\phi_1$ rather than~$\phi_0$, according to ${\delta_t\phi=G(t)-\gamma[\phi-\phi_1(t)]}$.

Figure~\ref{fg:nematic-fit}(a) compares the $C_4$-odd photomodulation response~$\delta\phi$ measured at~14~K, 16~K, and~18~K with fits obtained by integrating the rate equation that accounts for coupled order.  The quality of fit is excellent; in particular, the reversal of sign for~$t\sim\gamma^{-1}$ is accurately reproduced.  The fitted parameters are $\lambda'\equiv \lambda \chi_\phi(\phi^*_0/\phi_0)^2\delta|\psi(t)|^2$ and $A\equiv F\phi_0$, which are plotted as functions of temperature in Figs.~\ref{fg:nematic-fit}(b) and~(c), respectively.  Accurate fits are obtained with decay times for~$\phi$ and~$|\psi|^2$ fixed at 1.4~ps and 40~ps, respectively.  The physical picture of the coupled order parameter dynamics is as follows: immediately after photoexcitation, ${\Delta \phi(t)/\phi_0(T)=-F}$, just as in the normal state.  However, for times~$t\gg\gamma^{-1}$, a quasiequilibrium is reached in which $\phi(t)\approx \phi_1(t)$; that is, the time dependence of~$\phi(t)$ tracks~$\delta|\psi(t)|^2$.  Note that whereas~$\phi(t)$ is initially reduced relative to its equilibrium value, it is enhanced for $t\gg\gamma^{-1}$ due to the photoinduced suppression of superconducting order.

\section{\label{sec:region-transform}Aligning optical and Laue data}

\begin{figure}[h!]
  \centering
  \includegraphics[width=3.375in]{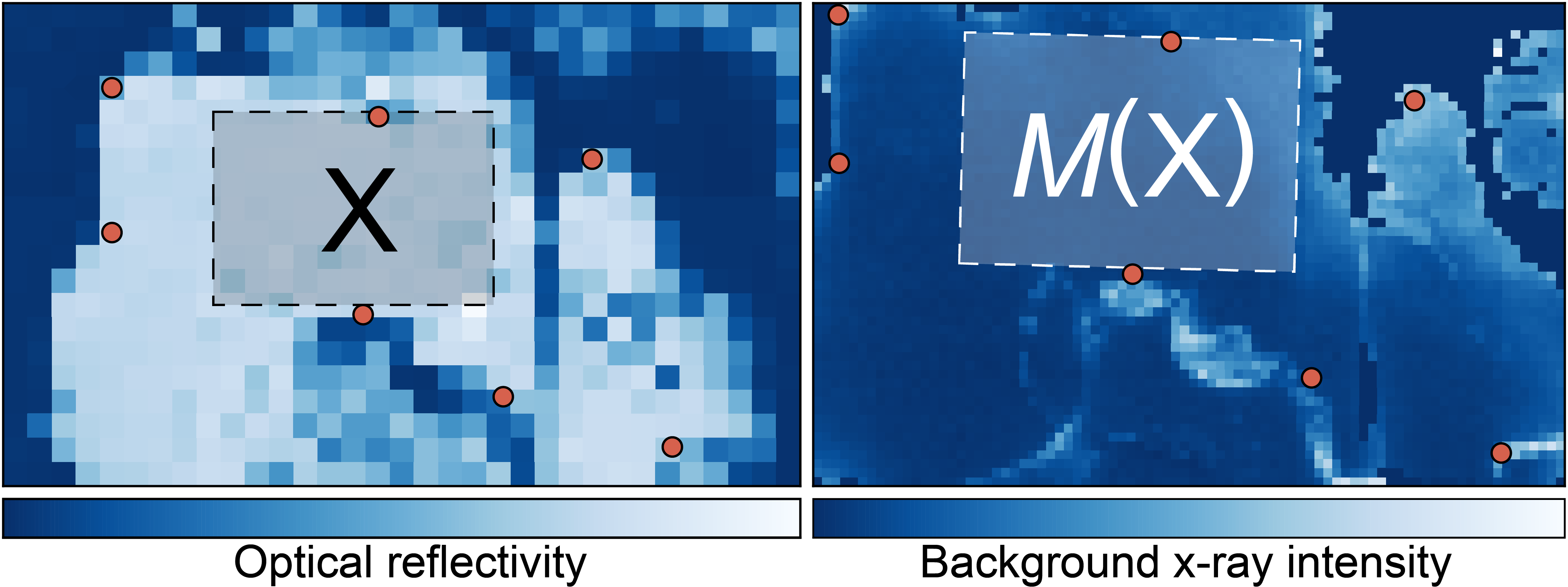}
  \caption{Large-scale maps of reflectivity (left) and background x-ray intensity (right) on an optimally doped P:Ba122 sample.  Shared features used as landmarks are marked by orange circles, and their coordinates are used to fit for a mapping~$M$ from the pump/probe region~$\text{X}$ to the corresponding subregion of the strain map,~$M(\text{X})$.
  \label{fg:landmarking}}
\end{figure}

The photomodulation and micro-Laue experiments were performed in different facilities and with slightly different sample orientations, so care was necessary to ensure that we compared the optical map with the correct subregion of the strain map.  Figure~\ref{fg:landmarking} shows large-scale images of reflectivity (left) and background x-ray intensity (right) on the P:Ba122 sample, with shared features marked by orange circles.  After landmarking the images in this fashion, we fit for a mapping~$M$ from the region corresponding to the photomodulation map,~$\text{X}$, to the corresponding region on the strain map,~$M(\text{X})$.  The alignment of various features shown in Fig.~\ref{fg:maps} indicates that this procedure was successful.

\newpage

\section{\label{sec:strain-free-mount}Strain is not caused by mounting or cooling}

\begin{figure}[h!]
  \centering
  \includegraphics[width=3.375in]{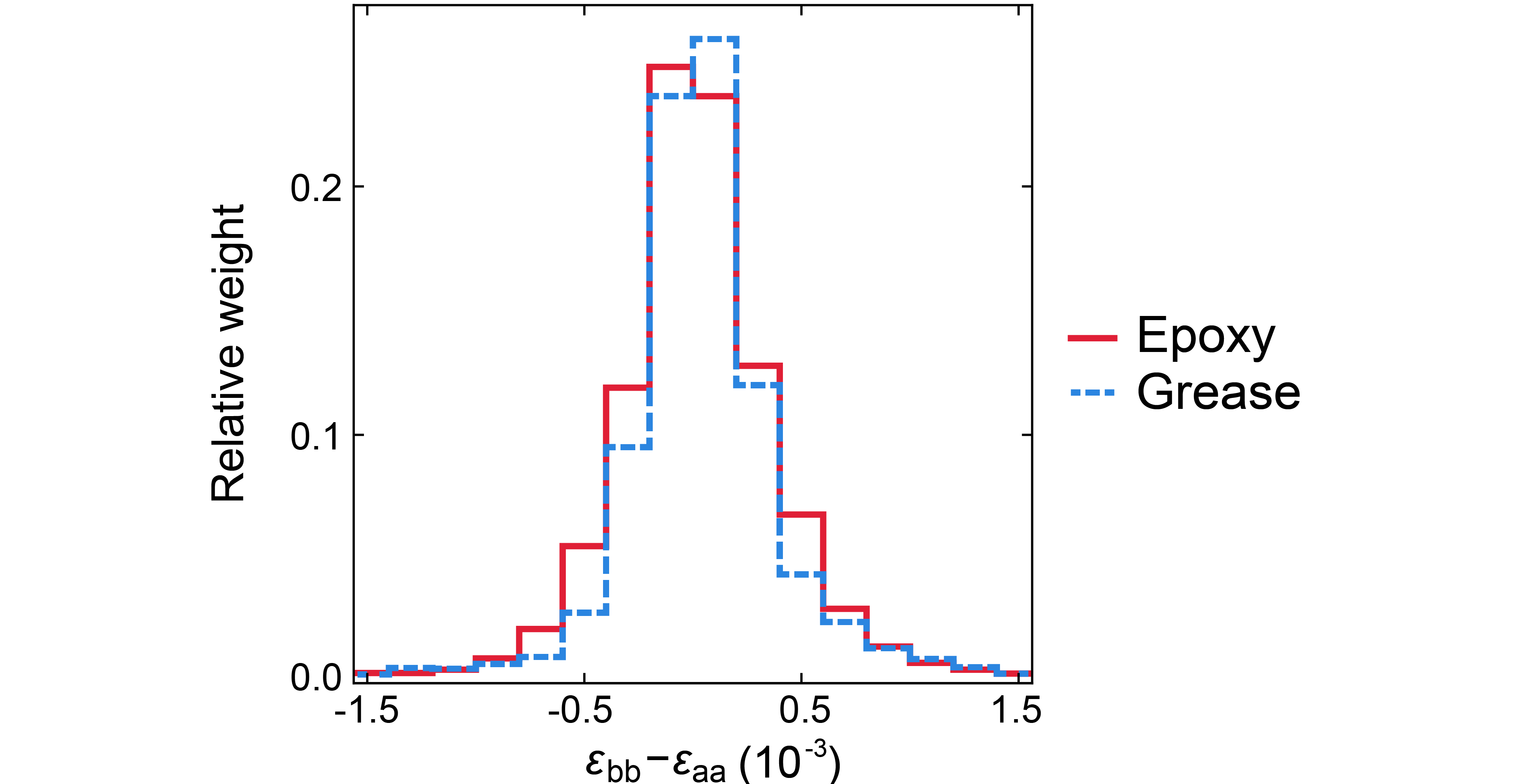}
  \caption{Distributions of \mbox{Fe$-$Fe} basis strain anisotropy in optimally doped P:Ba122 mounted with epoxy (solid red) and grease (dashed blue).
  \label{fg:strain-distribution}}
\end{figure}

In order to determine whether the strain variations we observed were intrisic or extrinsic, we performed micro-Laue measurements on a separate optimally doped P:Ba122 crystal mounted using grease and never cooled below room temperature.  Figure~\ref{fg:strain-distribution} shows histograms of the \mbox{Fe$-$Fe} basis strain anisotropy drawn from the original sample, mounted with epoxy, and the grease-mounted sample.  The strain anisotropy distributions are qualitatively very similar, indicating that the strain inhomogeneity we observe arises during sample growth.
\nopagebreak

\section{\label{sec:resistivity}Resistivity}

\begin{figure}[h!]
  \centering
  \includegraphics[width=3.375in]{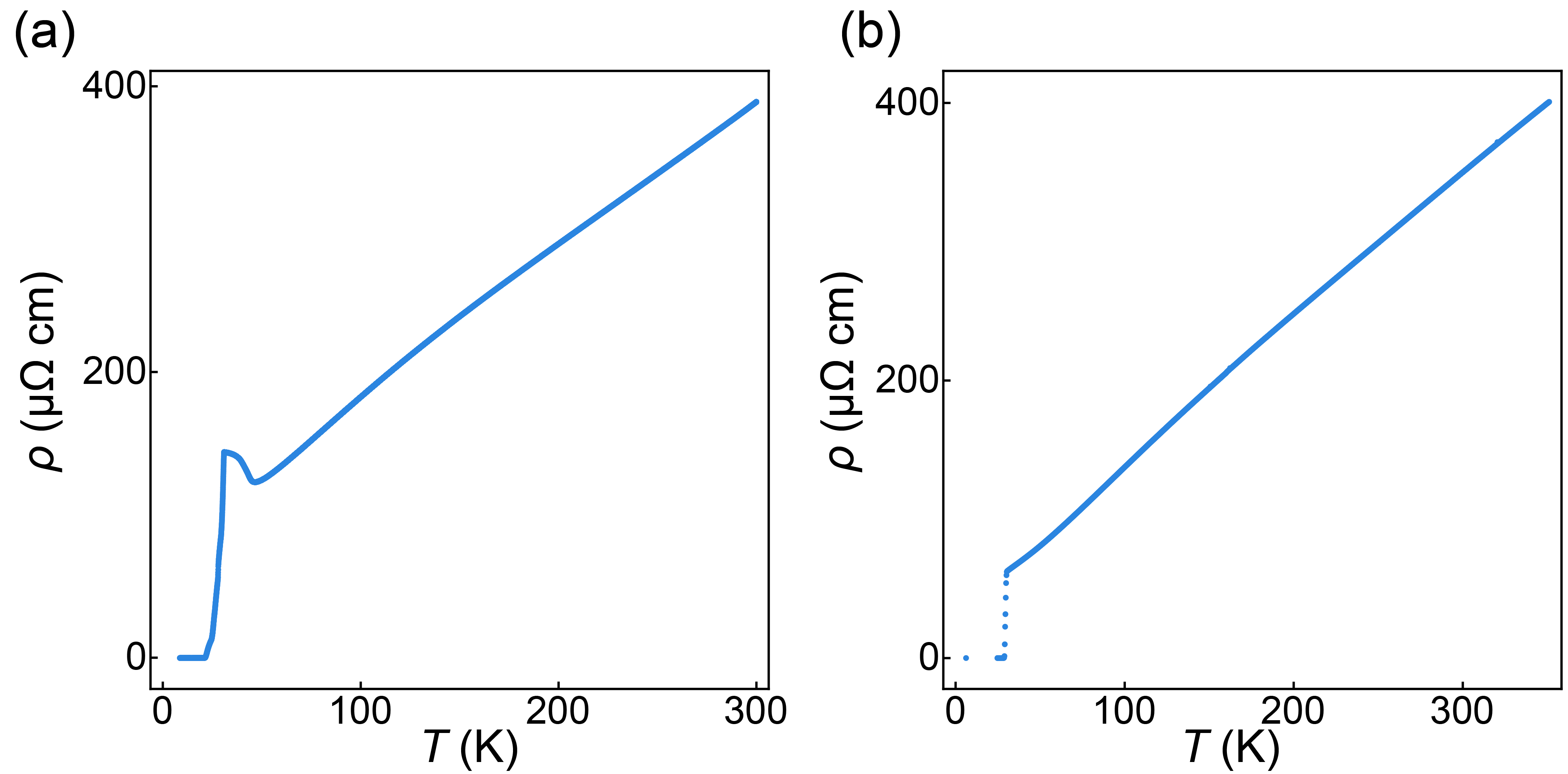}
    \caption{Resistivity as a function of temperature, measured on typical P:Ba122 samples with~(a)~${x=0.26}$ and~(b)~${x=0.31}$.
  \label{fg:resistivity}}
\end{figure}

We performed resistivity measurements to confirm sample quality.  Figures~\ref{fg:resistivity}(a-b) show typical temperature dependence curves for an underdoped and optimally doped sample, respectively.  In contrast to the significant resistivity increase as the sample cools through the N\'{e}el transition in the underdoped sample, the optimally doped sample has a monotonic temperature dependence and a sharp superconducting transition at~$30$~K.

\end{document}